\documentclass[%
 letterpaper,twocolumn,english,
 superscriptaddress,
showpacs,preprintnumbers,
 amsmath,amssymb,
 aps,
 prc,
floatfix,
]{revtex4-1}

\usepackage[usenames]{color}
\usepackage{amsmath}
\usepackage{amssymb}
\usepackage{longtable}  
\usepackage{graphicx}
\usepackage[dvipdfm,ps2pdf,unicode=true,bookmarks=false,breaklinks,pdfborder={0
    0 0},backref=false,colorlinks=true]{hyperref}  
\hypersetup{pdftitle={Single Spin Asymmetries in Charged Kaon Production from Semi-Inclusive Deep Inelastic Scattering on 
a Transversely Polarized $^3{\rm{He}}$ Target},pdfauthor={Y.X. Zhao , et. al. (The Jefferson Lab Hall
    A Collaboration)},pdfsubject={v12: nucl-ex,
    hep-ex},pdfkeywords={Transverse SSA SIDIS He-3 TMD Collins Sivers},linkcolor=DarkBlueCite, citecolor=DarkBlueCite,
  urlcolor=DarkBlueCite}  
\usepackage[hyphenbreaks]{breakurl}

\makeatletter

\usepackage{dcolumn}
\usepackage{bm}

\definecolor{DarkBlueCite}{rgb}{0.1,0.0,0.5}

\makeatother

\begin{document}



\author{Y.X.~Zhao}\email[Corresponding author: ]{yxzhao@jlab.org}
\affiliation{University of Science and Technology of China, Hefei
  230026, People's Republic of China}
\author{Y.~Wang}
\affiliation{University of Illinois, Urbana-Champaign, IL 61801}
\author{K.~Allada}
\affiliation{Massachusetts Institute of Technology, Cambridge, MA 02139}
\affiliation{Thomas Jefferson National Accelerator Facility, Newport
  News, VA 23606}  
\author{K.~Aniol}
\affiliation{California State University, Los Angeles, Los Angeles, CA 90032}
\author{J.R.M.~Annand}
\affiliation{University of Glasgow, Glasgow G12 8QQ, Scotland, United Kingdom}
\author{T.~Averett}
\affiliation{College of William and Mary, Williamsburg, VA 23187}
\author{F.~Benmokhtar}
\affiliation{Carnegie Mellon University, Pittsburgh, PA 15213}
\author{W.~Bertozzi}
\affiliation{Massachusetts Institute of Technology, Cambridge, MA 02139}
\author{P.C.~Bradshaw}
\affiliation{College of William and Mary, Williamsburg, VA 23187}
\author{P.~Bosted}
\affiliation{Thomas Jefferson National Accelerator Facility, Newport
  News, VA 23606}
% ...
\collaboration{The Jefferson Lab Hall A Collaboration}
\noaffiliation

\title{Single Spin Asymmetries in Charged Kaon Production from Semi-Inclusive Deep Inelastic Scattering on 
a Transversely Polarized $^3{\rm{He}}$ Target} 


\begin{abstract}
We report the first measurement of target single spin asymmetries 
of charged kaons produced in semi-inclusive deep inelastic 
scattering of electrons off a transversely polarized 
$^3{\rm{He}}$ target. Both the Collins and Sivers moments, which 
are related to the nucleon transversity and Sivers 
distributions, respectively, are extracted over the kinematic 
range of 0.1$<$$x_{bj}$$<$0.4 for $K^{+}$ and $K^{-}$ production. 
While the Collins and Sivers moments 
for $K^{+}$ are consistent with zero within the experimental 
uncertainties, both moments for $K^{-}$ favor negative 
values. The Sivers moments are compared to the theoretical 
prediction from a phenomenological fit to the world data. 
While the $K^{+}$ Sivers moments are consistent with the prediction, 
the $K^{-}$ results differ from the prediction at the 2-sigma level.
\end{abstract}


\maketitle

Significant progress has been made in recent years on our understanding 
of the transversity distribution as well as transverse-momentum-dependent 
parton distributions (TMDs) of the nucleons 
\cite{Barone2010267, J_Collins_TMD}.  
The nucleon transversity distribution \cite{Ralston1979109}, which represents the 
correlation between the quark transverse spin and the nucleon 
transverse spin, is suppressed in inclusive
deep inelastic scattering experiments due to its chiral-odd nature. 
While it was recognized that polarized Drell-Yan
experiments \cite{Ralston1979109, PAX_Drell_Yan} and 
Semi-Inclusive Deep Inelastic Scattering (SIDIS) experiments 
can both access the transversity distribution, our current 
knowledge on this distribution is mainly obtained from SIDIS.

The SIDIS processes, in which a hadron is detected
in coincidence with the scattered lepton 
\cite{BacchettaSIDIS, PhysRevD.71.034005,
Kotzinian1995234, Mulders1996197, 
PhysRevD.57.5780},
also involve another chiral-odd object, the so-called 
Collins fragmentation function \cite{Collins1993161}, to 
ensure helicity conservation. This allows the extraction 
of the transversity distribution, provided that the Collins 
fragmentation function is sizable. The Collins fragmentation 
functions were extracted to be significant by experiments at Belle \cite{PhysRevLett.96.232002}
and at BaBar \cite{BaBar_Collins}.

Pioneering efforts have been devoted towards the measurement 
of transversity distributions by the HERMES and COMPASS 
collaborations in dedicated SIDIS experiments using 
transversely polarized targets 
\cite{PhysRevLett.103.152002, Airapetian201011, Alekseev2010240}. 
A modulation of the form sin($\phi_{h}$ + $\phi_{S}$), 
the Collins moment, where $\phi_{h}$ and $\phi_{S}$ are the 
azimuthal angles of the detected hadron and the nucleon spin with 
respect to the lepton scattering plane, corresponds to a 
convolution of the transversity distribution and the Collins 
fragmentation function. Another important leading-twist 
TMD is the so-called Sivers function \cite{PhysRevD.41.83}, 
which represents the correlation between the nucleon transverse 
spin and the quark transverse momentum. It can be extracted 
through another angular modulation called the Sivers moment 
with the form of sin($\phi_{h}$ - $\phi_{S}$). Although the 
Sivers function is odd under the time reversal operation 
without exchanging the initial and final states 
\cite{Collins1993161}, it is allowed in the presence 
of QCD final-state interactions (FSI) between the 
outgoing quark and the target remnant
\cite{Brodsky200299, Ji2003383, Collins200243, Brodsky2002344}.

Results from the HERMES and COMPASS experiments have clearly 
shown the presence of the sin($\phi_{h}$ + $\phi_{S}$) and 
sin($\phi_{h}$ - $\phi_{S}$) modulations from proton 
targets \cite{PhysRevLett.103.152002, Airapetian201011, Alekseev2010240}. 
In remarkable contrast, much smaller modulations were found 
from deuteron targets \cite{Alekseev2009127}, suggesting 
that the process is flavor dependent. To shed new light 
on the flavor structure of the transversity and Sivers functions, 
it is important to extend SIDIS measurements to a polarized $^3{\rm{He}}$ target, 
whose spin comes predominantly from the neutron. 

The first such measurement was carried out on a polarized $^3{\rm{He}}$ 
target in Hall A at the Jefferson Laboratory and results for the charged 
pion SIDIS production have already been reported \cite{Qian2011, Huang2012}.
In this paper, we present the results on the azimuthal asymmetries in 
charged kaon SIDIS production. Since kaons contain strange quarks, 
the role of sea quarks in the nucleons with respect to the Collins 
and Sivers effects can be explored. The HERMES collaboration 
\cite{Airapetian201011} observed that the Collins effect from 
the proton target for $K^{+}$ is larger than that for $\pi^{+}$, 
while for $K^{-}$ the Collins effect is small and consistent with 
zero. They also reported that the Sivers effect for $K^{+}$ 
from the proton target is large and positive, but very small 
for $K^{-}$ \cite{PhysRevLett.103.152002}. 
The COMPASS collaboration reported that the Collins and Sivers 
effects for $K^{+}$ and $K^{-}$ production from the polarized 
deuteron target are consistent with zero \cite{Alekseev2009127}. 
Results from this work using a polarized $^3{\rm{He}}$ target 
will provide important new information to study the flavor 
dependent behavior of the Collins and Sivers effects.

The data were collected during experiment E06-010 at Jefferson Lab, Hall A. 
The experiment was conducted from November 2008 to February 2009 using 
a 5.9-GeV electron beam with an average current of 12 $\mu$A and a transversely 
polarized $^3{\rm{He}}$ target. Scattered electrons were detected in 
the BigBite spectrometer which was at $30^{o}$ to the beam right (facing 
the beam dump) with a momentum acceptance from 0.6 GeV/c to 2.5 GeV/c. 
Coincident charged hadrons ($\pi^{\pm}$, $K^{\pm}$ and protons) were detected 
in the High Resolution Spectrometer (HRS) \cite{Alcorn2004}, which was at $16^{o}$ 
to the beam left with a central momentum of 2.35 GeV/c. The electron beam helicity 
was flipped at a rate of 30 Hz. The unpolarized beam was achieved by summing the 
two helicity states, which differ by less than 100ppm per 1-hour run in beam charge.

The polarized $^3{\rm{He}}$ target consisted of a 40-cm long glass cell 
containing $\sim$10 atm of $^3{\rm{He}}$ and a small amount of $\text{N}_{2}$ 
to reduce depolarization \cite{Alcorn2004, Yiproc}. The ground state of $^3{\rm{He}}$ 
nuclear wavefunction is dominated by the S-state, in which the proton spins cancel 
each other and the nuclear spin is mostly carried by the neutron \cite{Bissey2001}.
Three pairs of Helmholtz coils were used in the experiment for producing the 
holding magnetic field in any direction. During the experiment, the target spin 
direction was oriented to transverse and vertical directions in order to 
enlarge the azimuthal angular coverage $\phi_{S}$. $^3{\rm{He}}$ nuclei were 
polarized by spin exchange optical pumping of a Rb-K mixture \cite{Babcock2005}.
Nuclear Magnetic Resonance (NMR) measurements, calibrated by the known water 
NMR signal and the electron paramagnetic resonance method, were performed 
to monitor the target polarization while the target spin direction was flipped 
every 20 minutes through adiabatic fast passage. An average in-beam target 
polarization of (55.4 $\pm$ 2.8)$\%$ was achieved during the experiment.

The BigBite spectrometer consisted of a single open dipole magnet, 
eighteen planes of multi-wire drift chambers organized in three groups 
and a scintillator plane sandwiched between lead-glass preshower and shower 
calorimeters. The magnetic field from the dipole, combined with tracking 
information from the drift chambers, was used to reconstruct the momenta of 
charged particles. Timing information for the scattered electrons was provided 
by the scintillators, and the electron trigger was formed by summing signals from 
two overlapping rows of preshower and shower blocks \cite{Xinthesis}.
The angular acceptance of the BigBite spectrometer was about 64 msr for a 40-cm target, 
which was essential to enlarge the 
azimuthal angular coverage $\phi_{h}$ for hadrons, given the small ($\sim$6 msr) 
angular acceptance of the HRS. A clean sample of electrons was
achieved by using two-dimensional cuts on the preshower energy $E_{\rm{ps}}$ and the 
momentum-dependent ratio $E/p$ in which $E$ and $p$ are the total energy deposit in the calorimeter
and the reconstructed momentum, respectively. After combining all the cuts, 
the $\pi^{-}$ contamination in the electron sample was less than 1\%.

\begin{figure}
\begin{center}
\includegraphics[width=85mm]{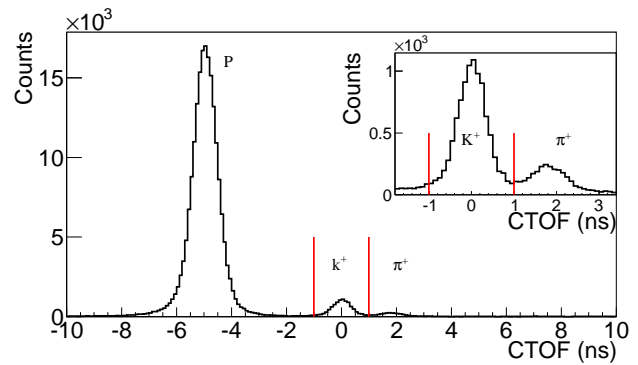}
\end{center}
\caption{(Color online) $^3{\rm{He}}$($e$, $e$'$h^{+}$)X coincidence timing spectrum after 
a cut on the aerogel detector to remove pions, 
where $h$ represents detected hadron. The kaon selection cuts are shown
as the two vertical lines. The top right sub-plot shows only 
$K^{+}$ and $\pi^{+}$ peaks in a relatively small CTOF range.} \label{tofplot}
\end{figure}

The HRS spectrometer configured for hadron detection consisted of two drift 
chambers for tracking, two scintillator planes for timing and triggering, 
a CO$_{2}$ gas Cerenkov detector and two layers of lead-glass calorimeter 
for electron rejection, an aerogel Cerenkov detector for pion rejection, 
and a ring imaging Cerenkov detector for hadron (pion, kaon, proton) 
identification \cite{youcai_thesis}. In addition, Coincidence Time Of Flight (CTOF)
between scattered electrons and hadrons was also recorded for hadron 
identification. Fig. \ref{tofplot} shows the CTOF spectrum. It 
describes the difference between the measured time of flight of the hadron 
and that of the expected kaon based on the electron timing. Therefore, 
the kaon peak is centered at zero and the proton, which is slower than the kaon, is 
peaked at a negative value. By applying a ``pion rejection" cut on the aerogel 
detector, pions were strongly suppressed, and the contamination of 
$\pi^{+}$ ($\pi^{-}$) in the $K^{+}$ ($K^{-}$) sample was less than 2\% (5\%). 
The random coincidence contamination in the $K^{+}$ ($K^{-}$) sample was 
less than 4\% (1\%), and the coincidental proton contamination in 
the $K^{+}$ sample was negligible.

\begin{figure}
\begin{center}
\includegraphics[width=85mm]{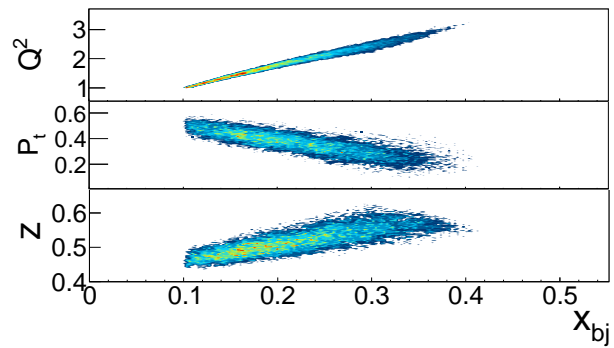}
\end{center}
\caption{(Color online) Correlation between $x_{bj}$ and kinematics 
variables ($Q^{2}$, $P_{t}$, $z$) for $K^{+}$, where 
$x_{bj}=\frac{Q^{2}}{2P \cdot q}$, 
$P_{t}=\sqrt{ {\vec{P_{h}}}^2 - ( \frac{ \vec{q} \cdot \vec{P_{h}} }{ |\vec{q}| }  )^2 }$,
$z=\frac{P \cdot P_{h}}{P \cdot q}$, $P$ is the four-momentum of the 
initial nucleon, $q$ is the four-momentum of the virtual photon,
$P_{h}$ is the four-momentum of the detected hadron.
} \label{kineplot}
\end{figure}

The SIDIS event sample for the analysis was selected by requiring: 
1) four-momentum-transfer squared $Q^{2}$ $>$ 1 GeV$^{2}$, 
2) virtual photon-nucleon invariant mass $W$ $>$ 2.3 GeV, 
3) the missing mass of undetected final-state particles $W'$ $>$ 1.6 GeV.
The kinematics coverage for $K^{+}$ is shown in Fig. \ref{kineplot}. 
After all the cuts, the total number of accepted SIDIS events were
about 10k and 2k for $K^{+}$ and $K^{-}$, respectively. The data were 
analyzed by using an azimuthally unbinned 
Maximum Likelihood Estimator (MLE) \cite{Jinthesis}. Due to the low statistics 
of the $K^{-}$ sample, the data were binned in one kinematical bin, while 
for $K^{+}$, the data were binned in four bins of $x_{bj}$. The central 
values for various kinematical variables are listed in Table \ref{kinetable}.
\begin{table}
\begin{center}
\begin{tabular}{|c|c|c|c|c|c|c|c|} \hline
       &$x_{bj}$&$y$&$z$&$Q^{2}$          &  $P_{t}$      &  $W$   &  $W'$ \\ 
       &	& & &  GeV$^2$        &  GeV          &  GeV   &   GeV \\  \hline   
$K^{+}$&0.137&0.85&0.48&1.29&0.46&3.0&2.08 \\ 
$K^{+}$&0.190&0.81&0.51&1.69&0.40&2.85&1.96 \\ 
$K^{+}$&0.250&0.77&0.53&2.11&0.33&2.69&1.83 \\ 
$K^{+}$&0.324&0.73&0.56&2.60&0.26&2.51&1.69 \\ \hline
$K^{-}$&0.210&0.80&0.51&1.83&0.38&2.80&1.93 \\ \hline
\end{tabular}
\caption{Tabulated central values for kinematical variables 
$x_{bj}$, $y$, $Q^{2}$, $z$, $P_{t}$, $W$, $W'$, where 
$y=\frac{q \cdot P}{l \cdot P}$, 
$W=\sqrt{ (P + q)^2 }$, 
$W'=\sqrt{(q + P - P_{h})^2}$, and $l$ is 
the four-momentum of the incoming lepton.
} \label{kinetable}
\end{center}
\end{table}

The likelihood was formed by the $\phi_{h}$ and $\phi_{S}$ dependent yield 
as shown in Eq. \eqref{MLEequation},
\begin{equation}
yield(\phi_{h}, \phi_{S}) 
= 
\rho \cdot \sigma \cdot a_{\pm}(\phi_{h}, \phi_{S})(1 + P\sum\limits_{j=1}^{2}\epsilon_{j}A_{j}(\phi_{h}, \phi_{S})) ,
\label{MLEequation}
\end{equation}
where $\rho$ is the target density, $\sigma$ is the unpolarized cross section, 
$a_{\pm}$($\phi_{h}$, $\phi_{S}$) is the acceptance for target spin state 
$\pm$, $A_{j}$($\phi_{h}$, $\phi_{S}$) is the $j^{th}$ azimuthal angular modulation, 
sin($\phi_{h}$ + $\phi_{S}$) or sin($\phi_{h}$ - $\phi_{S}$), $P$
is the target polarization, and $\epsilon_{j}$ is the amplitude of each modulation. 
The $\phi_{h}$ and $\phi_{S}$ definition 
follows the Trento Conventions \cite{PhysRevD.70.117504}.
The MLE method has been used for charged pion analysis \cite{Huang2012} and 
has been checked through Monte Carlo simulations. The results extracted from MLE
take into account the unbalanced beam charge associated with two target spin directions and 
the data acquisition livetime. The $^3{\rm{He}}$ Collins and 
Sivers moments were then obtained by correcting the dilution from unpolarized $\text{N}_2$ 
gas in the target cell. The nitrogen dilution factor is defined as
\begin{equation}
f_{\text{N}_{2}}\equiv\frac{\rho_{\text{N}_{2}}\sigma_{\text{N}_{2}}}{\rho_{^{3}\rm{He}}\sigma_{^{3}\text{He}}+\rho_{\text{N}_{2}}\sigma_{\text{N}_{2}}},
\label{dilequation}
\end{equation}
where $\rho$ is the density of the gas in the production target cell 
and $\sigma$ is the unpolarized SIDIS cross section. The ratio of 
unpolarized cross sections $\sigma_{\text{N}_{2}}$/$\sigma_{^3{\rm{He}}}$ 
was measured in dedicated runs on targets filled with
known amounts of unpolarized $\text{N}_{2}$ or $^3{\rm{He}}$ gas. 
The $f_{\text{N}_{2}}$ in this experiment was determined to be about 10$\%$.

The dominant systematic uncertainty in our measurement was the contamination from 
photon-induced charge-symmetric $e^{\pm}$ pairs, of which the $e^{-}$ was 
detected in BigBite. The yield of ($e^{+}$, $K^{\pm}$) coincidences was 
measured directly by reversing the magnetic field of BigBite, and hence the 
contamination of photon-induced electrons in the electron sample was determined.  
The contamination for $K^{-}$ detection was 14$\pm$7\%.
Hardly any events were observed in the latter 3 bins for $K^{+}$ detection 
from calibration runs which indicated that the contamination in these bins 
was small. To be conservative, the contaminations were given by a limit 
in these bins with the assumption that the contamination decreases linearly 
through 4 bins. The photon-induced electron contamination for $K^{+}$ was
determined to be 18.6$\pm$8.3\%, $<$10\%, $<$5\%, $<$3\%, respectively for 
the four $x_{bj}$-bins. Since this contamination is primarily from 
photon-induced pair production, it carries the same asymmetry as photon 
production. The asymmetry contamination correction for $K^{-}$ and the first bin 
of $K^{+}$ was given by the asymmetry from high energy 
$\gamma$-$K^{\pm}$ coincidence events. Additional experimental systematic 
uncertainties include: 
1) $\pi^{-}$ contamination in the electron sample, 
2) $\pi^{\pm}$ contamination in the $K^{\pm}$ sample, 
3) random coincidence contamination in the ($e^{-}$, $K^{\pm}$) coincidence sample, 
4) target density fluctuations, 
5) detector response drift caused by radiation damage to the BigBite calorimeter, 
6) target polarization, 
and 7) bin-centering effects. 
The quadrature sum of these uncertainties is quoted as the 
``experimental" systematic uncertainty for our measurement.

For the asymmetry extraction from Eq. \eqref{MLEequation}, we only 
included sin($\phi_{h}$ + $\phi_{S}$) and sin($\phi_{h}$ - $\phi_{S}$) 
modulations by neglecting other modulations, including 
sin(3$\phi_{h}$ - $\phi_{S}$) modulation at 
twist-2 \cite{Zhang:2013dow}, sin($\phi_S$) and 
sin(2$\phi_{h}$ - $\phi_{S}$) modulations at 
twist-3, Cahn cos($\phi_{h}$) and 
Boer-Mulders cos(2$\phi_{h}$) modulations from 
unpolarized cross section. The leakage from the longitudinal 
polarized target single spin asymmetry ($A_{UL}$) due 
to the small longitudinal component of the 
target polarization was also neglected. These effects were 
estimated by varying each term within an allowed range 
derived from the HERMES proton data \cite{Desy_phd_thesis}, 
assuming that the magnitude of each term for the neutron is 
similar to that of the proton. These effects were summed in 
quadrature to yield the ``fit" systematic uncertainty, which is 
dominated by the sin($\phi_{S}$) term.

\begin{figure}
\begin{center}
\includegraphics[width=85mm]{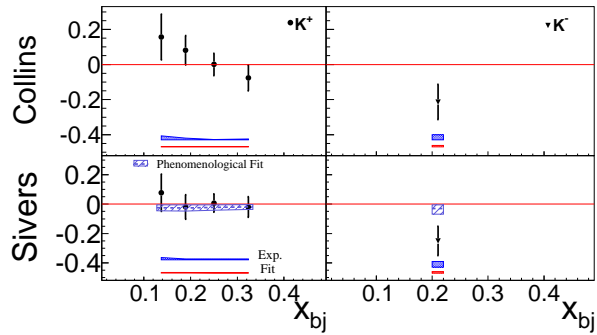}
\end{center}
\caption{(Color online) The extracted Collins and Sivers 
moments on $^3{\rm{He}}$ are 
shown together with their statistical errors and systematic error 
bands for both $K^{+}$ and $K^{-}$ electro-production.
The Sivers moments are compared to theoretical predictions from 
a phenomenological fit to the world data.} 
\label{he3resultplot}
\end{figure}

The extracted $^3{\rm{He}}$ Collins and Sivers moments are shown in 
Fig. \ref{he3resultplot} and tabulated in Table \ref{he3resulttable}. 
The error bars represent statistical uncertainties.
Experimental systematic uncertainties combined in quadrature from 
different sources are shown as a band labeled ``Exp.".
Systematic uncertainties due to neglecting other modulations are 
shown as a band labeled ``Fit".
The $K^{+}$ Collins and Sivers moments are consistent with zero within 
error bars, while for $K^{-}$ these moments
are found to favor negative values at the 2-sigma level.
\begin{table*}[h]
\begin{center}
\begin{tabular}{|c|c|c|c|} \hline
           &    $x_{bj}$    &     Collins moment                    & Sivers moment \\ \hline
$K^{+}$    &    0.137       &    0.16$\pm$0.13$\pm$0.024(0.003)     &  0.078$\pm$0.13$\pm$0.017(0.005) \\
$K^{+}$    &    0.190       &    0.082$\pm$0.083$\pm$0.01(0.002)    & -0.019$\pm$0.083$\pm$0.0065(0.004) \\ 
$K^{+}$    &    0.250       &    0.0009$\pm$0.063$\pm$0.003(0.002)  &  0.0074$\pm$0.063$\pm$0.006(0.003) \\ 
$K^{+}$    &    0.324       &    -0.075$\pm$0.074$\pm$0.006(0.002)  & -0.019$\pm$0.07$\pm$0.006(0.002) \\ \hline
$K^{-}$    &    0.210       &      -0.21$\pm$0.10$\pm$0.03(0.009)   & -0.25$\pm$0.10$\pm$0.039(0.01)  \\ \hline
\end{tabular}
\caption{Tabulated $^3{\rm{He}}$ results for the central kinematical variable $x_{bj}$. 
The format for the tabulated results follows ``central 
value" $\pm$ ``statistical uncertainty" $\pm$ ``experimental systematic 
uncertainty (systematic uncertainty due to fit model)".} \label{he3resulttable}
\end{center}
\end{table*}

The Sivers moments from the $^3{\rm{He}}$ target are compared to theoretical 
predictions from a phenomenological fit to the world data \cite{Anselmino:2008sga}.
While $K^{+}$ Sivers moments are consistent with the prediction, 
$K^{-}$ results differ from the prediction at the 2-sigma level.
Due to the lack of information on the Collins fragmentation function 
for kaons, no theoretical predictions on the Collins moments are currently
available. Our data on the Collins moments will provide independent inputs 
for a future global analysis to extract flavor dependent transversity distributions.
Although with large uncertainties for the $K^{-}$ Collins and Sivers moments, 
the results are still surprising compared to our current knowledge of
the effects of sea quarks and unfavored fragmentation functions. 
Therefore, to fully understand the sea quark flavor dependence of 
the Collins and Sivers moments, high-precision kaon data are required for 
transversely polarized proton, deuteron and $^3{\rm{He}}$ targets.

In summary, we have reported the first measurement of target 
single spin asymmetries of charged kaons produced in 
SIDIS using a transversely polarized $^3{\rm{He}}$ target. 
Our data show that the Collins and Sivers moments for $K^{+}$ are consistent 
with zero within the experimental uncertainties,  
while the $K^{-}$ results favor negative values. 
While the statistics for the 6-GeV E06-010 measurements were limited, 
experiment E06-010 laid the foundation for
future 12-GeV SIDIS experiments at JLab \cite{Jlab12GeV}. 
These future SIDIS experiments will provide us a unique opportunity in mapping
the kaon Collins and Sivers moments to much higher precision.

We acknowledge the outstanding support of the JLab Hall A technical 
staff and the Accelerator Division in accomplishing this experiment. 
This work was supported in part by the U. S. National Science Foundation, 
and by DOE contract number DE-AC05-06OR23177, under which the Jefferson 
Science Associates (JSA) operates the Thomas Jefferson National Accelerator Facility.

\nocite{*}
\bibliography{references}
\end{document}